# Chiral coupling between magnetic layers with orthogonal magnetization

Can Onur Avci, Charles-Henri Lambert, Giacomo Sala and Pietro Gambardella

*Department of Materials, ETH Zürich, CH-8093 Zürich, Switzerland*

**We report on the occurrence of strong interlayer Dzyaloshinskii-Moriya interaction (DMI) between an in-plane magnetized Co layer and a perpendicularly magnetized TbFe layer through a Pt spacer. The DMI causes a chiral coupling that favors one-handed orthogonal magnetic configurations of Co and TbFe, which we reveal through Hall effect and magnetoresistance measurements. The DMI coupling mediated by Pt causes effective magnetic fields on either layer of up to 10-15 mT, which decrease monotonously with increasing Pt thickness. Ru, Ta, and Ti spacers mediate a significantly smaller coupling compared to Pt, highlighting the essential role of Pt in inducing the interlayer DMI. These results are relevant to understand and maximize the interlayer coupling induced by the DMI as well as to design spintronic devices with chiral spin textures.**

The ability to engineer the coupling between magnetic layers is central to reveal emergent magnetic and electronic interactions at interfaces as well as to improve the functionality of magnetic sensors, non-volatile memories, and logic gates. Magnetic layers can couple directly through short-range exchange interactions when they are in contact with each other, or indirectly through a nonmagnetic (NM) spacer. A prominent manifestation of direct coupling is the exchange bias between adjacent ferromagnetic (FM) and antiferromagnetic layers, which allows for tuning the hysteretic behavior of the FM layers [1,2]. Apart from dipolar coupling [3,4], the most studied type of indirect coupling is the Ruderman–Kittel–Kasuya–Yosida (RKKY) interaction [5-10] between two FM layers mediated by the conduction electrons of a NM spacer [8,11,12]. This coupling has an oscillatory nature that favors either parallel or antiparallel alignment of the magnetization of the FMs depending on the thickness of the NM spacer. The theoretical and material aspects of the RKKY-driven interlayer coupling are well understood in conventional FM/NM/FM trilayers, where NM is usually Cu, Cr or Ru [12,13].

Recently, increasing attention has been devoted to the coupling mediated by the interfacial Dzyaloshinskii-Moriya interaction (DMI) [14-20]. The DMI is an antisymmetric exchange interaction that favors the orthogonal alignment of neighboring spins in materials with spatial inversion asymmetry. The DMI was originally investigated in bulk systems such as $\alpha$-$Fe_2O_3$ and the B20 compounds [21]. However, theoretical work has shown that a strong DMI emerges at FM/NM interfaces with broken inversion symmetry and strong spin-orbit coupling [22-26], which stabilizes chiral spin textures such as Néel domain walls (Fig. 1 – left panels), and skyrmions [14-16,27-38]. The DMI in these systems stems from an additional term in the RKKY interaction due to spin-orbit scattering of the conduction electrons by the atoms of the NM layer, as exemplified by the three-site model of Levy and Fert [39,40]. Atomistic Monte Carlo calculations have shown that this model can be extended to FM/NM/FM trilayers (Fig. 1 – right panels), where the DMI promotes nontrivial three-dimensional spin textures with both intralayer and interlayer chiralities [41]. The interlayer coupling mediated by the DMI thus offers novel opportunities to tune the magnetic texture and functionality of magnetic multilayers.

The occurrence of interlayer DMI was recently demonstrated in FM/Pt/Ru/Pt/FM multilayers with parallel or antiparallel magnetization, in which Pt promotes the perpendicular magnetic anisotropy and DMI of the FMs, and Ru mediates the RKKY coupling between them [17,18]. In such systems, the DMI results in canted magnetic structures with chiral exchange-biased hysteresis loops. In this Letter, we demonstrate strong chiral coupling between an out-of-plane (OOP) ferrimagnet, TbFe, and an in-plane (IP) FM, Co, through a single Pt spacer layer. This orthogonal configuration maximizes the DMI between layers, leading to significant changes of the coercivity depending on the relative orientation of the OOP and IP magnetizations. We report effective DMI fields ($B_{DMI}$) of up to 13 mT, which are significantly larger than those found in previous work. We further devise an experimental procedure to independently quantify $B_{DMI}$ for the IP and OOP layers, and to compare different spacer materials without changing the FMs. We demonstrate that Pt is a key material for mediating and inducing the interlayer DMI, in analogy with the interface-driven DMI in FM/NM bilayers. Moreover, we show that the coupling decreases monotonically with Pt thickness and becomes weaker if Pt is replaced by Ru, Ta, or Ti.

Our samples are //Ti(3)/TbFe(8)/Co(0.4)/X($t$)/Co(3)/Ti(5) layers deposited by d.c magnetron sputtering onto a Si/SiO$_2$ substrate at room temperature [Fig. 2(a)]. The numbers correspond to the thickness in nm. X is the spacer layer Pt, Ti, Ru or Ta with thickness $t$. For Pt, $t$ was varied between 1.0 and 3.0 nm, for all other elements $t$ was fixed to 1.7 nm. All layers were grown in a base pressure of ~$5 \times 10^{-8}$ mbar and Ar partial pressure of $2 \times 10^{-3}$ mbar. The composition of TbFe was 35% Tb and 65% Fe, optimized to have bulk OOP anisotropy. The top Co(3) layer has IP anisotropy in all samples. An ultrathin Co(0.4) was deposited between TbFe and the spacer layer in order to enhance the magnetic coupling between TbFe and the top Co layer, as indirect couplings are stronger between $d$-electron systems. This layer is assumed to be magnetically coupled to the Fe sublattice of TbFe and will not be mentioned explicitly in the remainder of this paper. We used photolithography and lift-off to fabricate 5 μm-wide Hall bars, as shown in Fig. 2(b). We measured the anomalous Hall effect and magnetoresistance at room temperature using a low amplitude a.c. current and the harmonic detection method [42] in order to probe the magnetization of the Co and TbFe layers.

We first focus on the samples with Pt spacers. Figure 2(c) shows a typical Hall resistance ($R_H$) measurement as a function of OOP magnetic field for $t$ = 1.2 nm. In this geometry, $R_H$ is proportional to the OOP magnetization due to the anomalous Hall effect, which results in the superposition of signals from both the TbFe and Co layers. We identify the OOP magnetization of TbFe ($\boldsymbol{M}_{TbFe}$) with the sharp reversal and hysteretic behavior around $B$ = 0, and the magnetization of Co ($\boldsymbol{M}_{Co}$) with the gradual increase of $R_H$ up to the saturation field $B$ ~ 0.6 T, which is due to the rotation of the Co magnetic moments from IP to OOP with increasing field. Qualitatively, $R_H$ can be decomposed into a square-like hysteresis loop attributed to TbFe and a hard axis loop attributed to Co, as shown in Fig. 2(d). The positive $R_H$ for the OOP component indicates that $\boldsymbol{M}_{TbFe}$ is parallel to that of the Fe sublattice [43]. Measurements of $R_H$ in samples with different spacer layers display a similar behavior. These measurements unequivocally show that the magnetizations of the two layers are orthogonal to each other at equilibrium, which maximizes the DMI coupling.

We now describe the expected behavior in orthogonally-oriented layers resulting from the effective Hamiltonian $\mathcal{H}_{DMI} = -\boldsymbol{D} \cdot \boldsymbol{M}_1 \times \boldsymbol{M}_2$, where $\boldsymbol{D}$ is the DMI vector and $\boldsymbol{M}_{1,2}$ the magnetization of each layer. In asymmetric trilayers, $\boldsymbol{D}$ is constrained by symmetry to lie in the *xy* plane [40,41,44]. Thus, the interlayer DMI favors a unique sense of rotation of $\boldsymbol{M}_1$ and $\boldsymbol{M}_2$ in the plane orthogonal to $\boldsymbol{D}$. Unlike for Néel domain walls in a single FM layer, however, the handedness of the chirality cannot be defined in a unique way in a FM/NM/FM trilayer. Moreover, the in-plane direction of $\boldsymbol{D}$ is not defined a priori in a multilayer with close-packed stacking [41]. For the sake of the discussion, we assume $\boldsymbol{D} \parallel -\boldsymbol{y}$, $\boldsymbol{M}_1 \equiv \boldsymbol{M}_{\text{TbFe}}$ and $\boldsymbol{M}_2 \equiv \boldsymbol{M}_{\text{Co}}$, which gives a chirality that favors the configurations ($\boldsymbol{M}_{\text{TbFe}} \parallel \boldsymbol{z}$, $\boldsymbol{M}_{\text{Co}} \parallel -\boldsymbol{x}$) and ($\boldsymbol{M}_{\text{TbFe}} \parallel -\boldsymbol{z}$, $\boldsymbol{M}_{\text{Co}} \parallel \boldsymbol{x}$) over the opposite ones ($\boldsymbol{M}_{\text{TbFe}} \parallel \boldsymbol{z}$, $\boldsymbol{M}_{\text{Co}} \parallel \boldsymbol{x}$) and ($\boldsymbol{M}_{\text{TbFe}} \parallel -\boldsymbol{z}$, $\boldsymbol{M}_{\text{Co}} \parallel -\boldsymbol{x}$). The macroscopic manifestation of such a coupling is an effective magnetic field $\boldsymbol{B}_{\text{DMI}} = \boldsymbol{D} \times \boldsymbol{M}_{\text{TbFe}}$ acting on $\boldsymbol{M}_{\text{Co}}$ and directed along -*x* (+*x*) when $\boldsymbol{M}_{\text{TbFe}}$ points along +*z* (-*z*). Likewise, a field $\boldsymbol{B}_{\text{DMI}} = -\boldsymbol{D} \times \boldsymbol{M}_{\text{Co}}$ will act on $\boldsymbol{M}_{\text{TbFe}}$ and pull it along +*z* (-*z*) when $\boldsymbol{M}_{\text{Co}}$ is oriented along -*x* (+*x*). We tested this hypothesis by examining the field-induced magnetization reversal behavior of TbFe/NM/Co trilayers in different experimental geometries.

We characterized $B_{\text{DMI}}$ acting on the OOP layer by sweeping the external magnetic field $\boldsymbol{B}$ at an oblique angle $\theta_B$ relative to $\boldsymbol{z}$ and with an in-plane projection parallel to $\pm \boldsymbol{D} \times \boldsymbol{M}_{\text{TbFe}}$ [Fig. 3(a) – top diagrams]. Starting from $\boldsymbol{M}_{\text{TbFe}} \parallel \boldsymbol{z}$, the field sweep with $\boldsymbol{B}$ tilted towards $\boldsymbol{D} \times \boldsymbol{z}$ will favor the DMI-stabilized configurations, leading to a reduction of the coercive field $B_c$, whereas the sweep with $\boldsymbol{B}$ tilted towards $-\boldsymbol{D} \times \boldsymbol{z}$ will force unfavorable magnetic configurations and increase $B_c$ (see Supplemental Material). Since $\boldsymbol{D}$ is not known a priori, we performed field sweep measurements on TbFe/Pt(1.5 nm)/Co trilayers for different angles $\varphi_B$ of the IP component of $\boldsymbol{B}$ relative to the *x* direction with $\theta_B$ fixed at 15°. Indeed, we observed a clear difference of $B_c$ in the hysteretic loop of TbFe depending on $\varphi_B$, which cannot be associated to field misalignment (see Supplemental Material). Figure 3(b) shows the data taken with the field initially tilted towards $\varphi_B = 135°$ and 315°, where we obtained the maximum difference in $B_c$ between the two measurements described above. The coercivity difference is calculated as $\Delta B_c(\varphi_B, B > 0) = B_{c1} - B_{c3}$ and $\Delta B_c(\varphi_B + \pi, B < 0) = B_{c2} - B_{c4}$ [$B_{c1}$, $B_{c2}$, $B_{c3}$ and $B_{c4}$ are defined in Fig. 3(b)] and plotted in Fig. 3(c) for different angles $\varphi_B$. We find that $\Delta B_c$ varies as a sine function, which is the expected behavior since the $M_{Co}$ follows the IP component of $B$ and $B_{DMI}$ is expected to scale proportionally to the projection of $M_{Co}$ on $D \times z$. The sinusoidal fit $\Delta B_c = 2B_{DMI} \sin(\varphi_B - \varphi_0) / \cos \theta_B$ gives $\varphi_0 = 45°$ as the $D$ direction and gives $B_{DMI} = 8.1 \pm 0.3$ mT for this dataset. These measurements unequivocally demonstrate the influence of strong interlayer DMI coupling on $M_{TbFe}$. Other forms of coupling favoring collinear alignment of $M_{Co}$ and $M_{TbFe}$, such as proximity-mediated ferromagnetic coupling, RKKY and dipolar coupling, are excluded because they would not lead to asymmetric magnetization curves relative to $\pm \theta_B$ and a sinusoidal variation of $\Delta B_c$. An alternative experimental scheme to measure $B_{DMI}$ by rotating the magnetic field about $D$ is described in the Supplementary Material.

Next, we measured the influence of the interlayer DMI on the IP layer. As mentioned above, $B_{DMI}$ is expected to act on $M_{Co}$ as an IP bias field pointing towards $D \times (\pm z)$ depending on the positive or negative orientation of $M_{TbFe}$. We verify this by probing the magnetoresistance as a function of IP magnetic field. Figure 3(d) shows the change of the longitudinal resistance $R$ during an IP field sweep applied at $\varphi_B = 0°$ for $M_{TbFe}$ pointing up and down. In this case, we show the data acquired when the positive field direction makes an angle of 45° with respect to $+D \times z$ because this angle increases the excursion of the magnetoresistance due to the inversion of the magnetization. We observe that, when $M_{TbFe}$ is up (down), $R$ has a minimum at a negative (positive) applied field. Whereas the minimum in the magnetoresistance is a signature of domain formation around the reversal field, the shift $\Delta B_s$ between the two curves indicates that a net bias field acts on $M_{Co}$. As the direction of the IP bias field depends on the sign of $M_{TbFe}$, we associate $\Delta B_s$ with $B_{DMI}$ acting on the Co layer, giving $B_{DMI} = \Delta B_s/2$. Macrospin simulations including the interlayer DMI in addition to the Zeeman and magnetic anisotropy energy are in excellent agreement with the data reported in Fig. 3 for both OOP and IP field sweeps, thus supporting our interpretation of the data (see Supplemental Material).

We then quantified the Pt thickness dependence of the interlayer DMI by performing a full set of measurements for each TbFe/Pt($t$)/Co sample as described in Fig. 3. Figures 4(a) and (b) show $B_{DMI}$ as a function of $t$ measured on the TbFe and Co layers, respectively. In both cases, $B_{DMI}$ exceeds 10 mT for $t < 1.5$ nm, and decreases with increasing spacer thickness until it nearly vanishes at $t = 3$ nm. For $B_{\text{DMI}}$ acting on $\boldsymbol{M}_{\text{TbFe}}$ we observe small deviations for Pt(1) and Pt(2) out of the overall monotonic decreasing trend. For $B_{\text{DMI}}$ acting on $\boldsymbol{M}_{\text{Co}}$, the decreasing trend is smoother, although the Pt(1) and Pt(2.5) data points are missing because the magnetoresistance measurements did not yield reproducible minima to estimate the bias field. Deviations in the DMI measured on the TbFe layer are attributed to variations in the structural and magnetic properties of these two samples with respect to the rest of the batch, as suggested by their different coercivity (see Supplemental Material). Overall, the data in Fig. 4(a) and (b) indicate that the interlayer DMI decays in a quasi-monotonic fashion with spacer thickness and survives up to $t \sim 3$ nm.

This thickness dependence is in contrast with the RKKY coupling mediated by Pt, which shows oscillations with a period of $t \sim 1\text{-}2$ nm in Pt/Co multilayers that are superimposed on a decreasing trend [45,46]. Moreover, the monotonic dependence on thickness cannot be explained by dipolar (orange peel) coupling, which would favor a collinear orientation of $\boldsymbol{M}_{\text{TbFe}}$ and $\boldsymbol{M}_{\text{Co}}$ [4]. In the three-site model [40,41], the DMI coupling between two magnetic atoms is a damped oscillatory function of the distance between them and the third nonmagnetic atom. In a magnetic multilayer, however, the total DMI is given by the sum of all the three-site interactions, including first and second nearest-neighbor nonmagnetic atoms [41], which might average out the oscillations as a function of thickness. The inevitable presence of crystalline defects and roughness is another possible cause for the monotonic damping of $B_{\text{DMI}}$. Interestingly, we also find that the direction of $\boldsymbol{D}$ differs between devices and samples with no specific trend. This is ascribed to the polycrystalline nature of our samples. The presence of ancillary uniaxial magnetic anisotropy in the Co layer might also influence the direction of $\boldsymbol{D}$ by setting a preferred direction for $\boldsymbol{M}_{\text{Co}}$.

Finally, we compare the interlayer DMI in systems with Pt, Ti, Ru and Ta spacers for a fixed thickness of 1.7 nm. Here the choice of elements allows us to compare Pt with light (Ti) and heavy elements (Ta) and with a strong RKKY-mediating material such as Ru. Figure 5 shows $\Delta B_c$ measured by performing field sweeps at different angles $\varphi_B$, similar to the measurements reported in Fig. 3b for Pt. We observe that all of the spacers except Pt show very small coupling. The sinusoidal fits of $\Delta B_c$ yield $B_{\text{DMI}} = 0.5 \pm 1.3$ mT for Ti, $2.6 \pm 0.5$ mT for Ru, and $2.9 \pm 0.5$ mT for Ta. These values are five to ten times smaller than $B_{\text{DMI}}$ observed for Pt. We attribute the smaller DMI observed in Ti and Ru to their smaller spin-orbit coupling. Ta, despite its large spin-orbit coupling, is known to generate much smaller interfacial DMI with respect to Pt [30,47], in agreement with our findings. This set of measurements shows that FM/NM/FM trilayers with OOP-IP magnetization can be used to probe the interlayer DMI in a wide range of NM materials.

There is no extensive literature on the interlayer DMI to compare the above results, except the experiments on FM/Pt/Ru/Pt/FM multilayers reported in Refs. [17,18]. In these systems, the maximum $B_{\text{DMI}}$ is 4 mT, which is significantly smaller than $B_{\text{DMI}}$ in our TbFe/Pt/Co trilayers, despite the larger FM thickness in our system. As the interlayer coupling is mediated by the FM/NM interfaces, we expect $B_{\text{DMI}}$ to decrease when the volume of the FM increases. Therefore, the higher $B_{\text{DMI}}$ reported here cannot be ascribed to the different thickness of the FM, but rather to the use of Pt instead of Pt/Ru/Pt as a coupling layer and to the orthogonal OOP-IP magnetic configuration of the FM layers.

We can alternatively compare our findings with the interfacial intralayer DMI of FM/NM bilayers. For a quantitative comparison we convert the maximum $B_{\text{DMI}}$ found for the Pt(1.2) spacer into an interfacial DMI energy $E_{\text{DMI}} = B_{\text{DMI}} M_s t_{\text{FM}}$. Here $M_s$ and $t_{\text{FM}}$ are the saturation magnetization and the thickness of the ferromagnetic layer, respectively, on which $B_{DMI}$ is acting. By using $M_s^{\text{Co}} = 1.2 \times 10^6$ A/m and $M_s^{\text{TbFe}} = 0.4 \times 10^6$ A/m obtained from similar layers we find $E_{\text{DMI}}^{\text{Pt/Co}} = 43$ µJ/m² and $E_{\text{DMI}}^{TbFe/Pt} = 44$ µJ/m². The quantitative agreement between the two $E_{\text{DMI}}$ values stems from the reciprocity of the effect on the opposite interfaces of Pt. The interlayer $E_{\text{DMI}}$ is thus significantly smaller than the intralayer $E_{\text{DMI}} \sim 1$ mJ/m² reported for Pt/FM interfaces [34,35,47-50]. The smaller interlayer DMI correlate with the damping of $B_{\text{DMI}}$ as a function of Pt thickness. Thinner Pt spacer layer could, in principle, generate stronger interlayer DMI. However, direct ferromagnetic coupling between the FM layers mediated by the proximity effect in Pt would tilt $\boldsymbol{M}_{\text{Co}}$ OOP, thus reducing the angle between $\boldsymbol{M}_{\text{TbFe}}$ and $\boldsymbol{M}_{\text{Co}}$ and likewise the DMI. Finally, we note that the interlayer DMI can coexist with proximity coupling and other types of collinear couplings, but the latter are nonzero only if $\boldsymbol{M}_1 \cdot \boldsymbol{M}_2 \neq 0$.

In conclusion, we demonstrated strong interlayer DMI coupling in FM/NM/FM trilayers with orthogonal magnetization and NM = Pt. Our coupled IP-OOP stack is optimized to give rise to maximum DMI, which leads to an effective field $B_{\mathrm{DMI}} \approx 13$ mT for $t = 1.2$ nm, corresponding to $E_{DMI} \approx 44$ μJ/m$^2$. We show that the DMI coupling decreases monotonically with increasing $t$ and vanishes at $t = 3$ nm. The samples with Ti, Ta, and Ru spacers show significantly lower interlayer DMI with respect to Pt, in qualitative agreement with the lower intralayer DMI induced by these materials on single FM layers. Our experimental scheme allows for quantifying the interlayer DMI acting on either one of the two FM layers using simple Hall effect and magnetoresistance measurements, and is readily applicable to a variety of experimental systems. These results provide insight into novel mechanisms for tuning the coupling between magnetic layers. Such a strong DMI coupling could be harnessed to design vertically-stacked heterostructures with correlated magnetization for use in logic and memory spintronic devices.

**Acknowledgments**

We acknowledge support by the Swiss National Science Foundation through grants #200020_172775 and PZ00P2-179944.

**FIGURES**

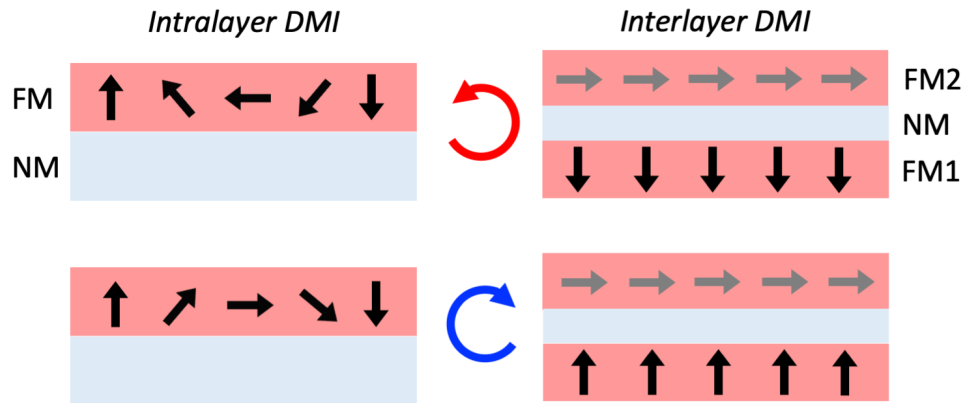

**Figure 1** – Schematic representation of the interfacial DMI coupling illustrating the intralayer (left) and interlayer (right) coupling scenarios. The black and gray arrows represent local magnetization in the ferromagnetic (FM) layers with OOP and IP magnetic anisotropy, respectively.

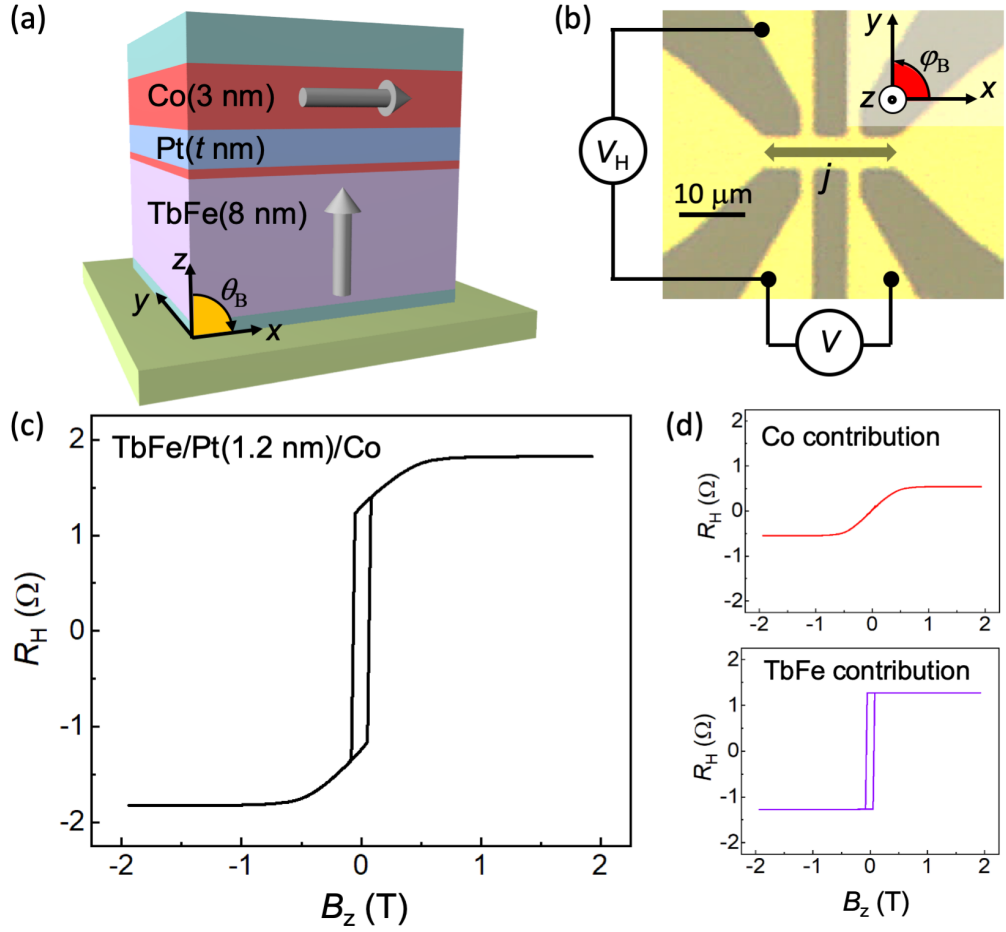

**Figure 2** – (a) Sketch of the multilayer structure (only relevant layers are labeled) and coordinate system. The block arrows indicate the magnetization of the top and bottom layers. (b) Device micrograph, electrical connections, and coordinate system. $j$ is the current density, $\varphi_B$ is the in-plane field angle (c) Hall resistance of TbFe(8)/Co(0.4)/Pt(1.2)/Co(3) during a sweep of the OOP field ($B_z$). (d) Separation of $R_H$ due to the top Co layer ($\boldsymbol{M}_{Co}$) and bottom TbFe/Co layer ($\boldsymbol{M}_{TbFe}$) obtained from the data shown in (c) by assuming a linear field dependence for $\boldsymbol{M}_{Co}$ between ±0.3 T and constant $R_H$ for $\boldsymbol{M}_{TbFe}$ outside the coercivity region.

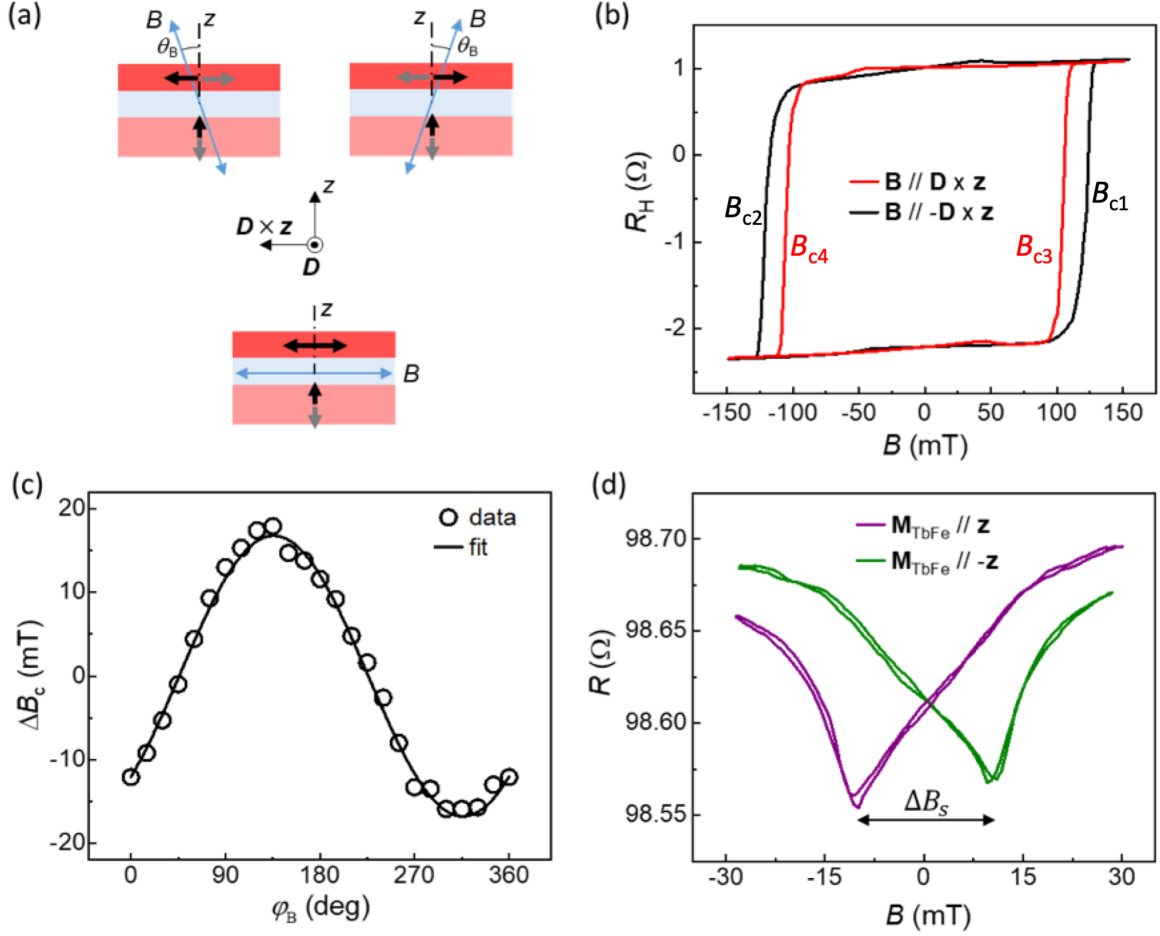

**Figure 3** – (a) Schematics of the field sweep measurements with tilt angle favoring and opposing the interlayer DMI (top left and right, respectively). The bottom diagram shows the geometry employed for the magnetoresistance measurements. (b) Hall resistance of TbFe(8)/Co(0.4)/Pt(1.5)/Co(3) measured during a field sweep at $\theta_B = 15°$ tilted along $\mathbf{D} \times \mathbf{z}$ (red line, initial $\varphi_B = 135°$) and along $-\mathbf{D} \times \mathbf{z}$ (black line, initial $\varphi_B = 315°$). (c) Coercivity difference $\Delta B_c$ as a function of $\varphi_B$. The line is a fit to the data (see text). (d) Magnetoresistance as a function of IP field for two different orientations of $\mathbf{M}_{TbFe}$. The IP field is applied at $\varphi_B = 0°$ such that the positive field direction has a positive projection on $-\mathbf{D} \times \mathbf{z}$. $\Delta B_s$ is the difference between the IP switching fields of $\mathbf{M}_{Co}$ that is used to quantify $B_{DMI}$ acting on the Co layer.

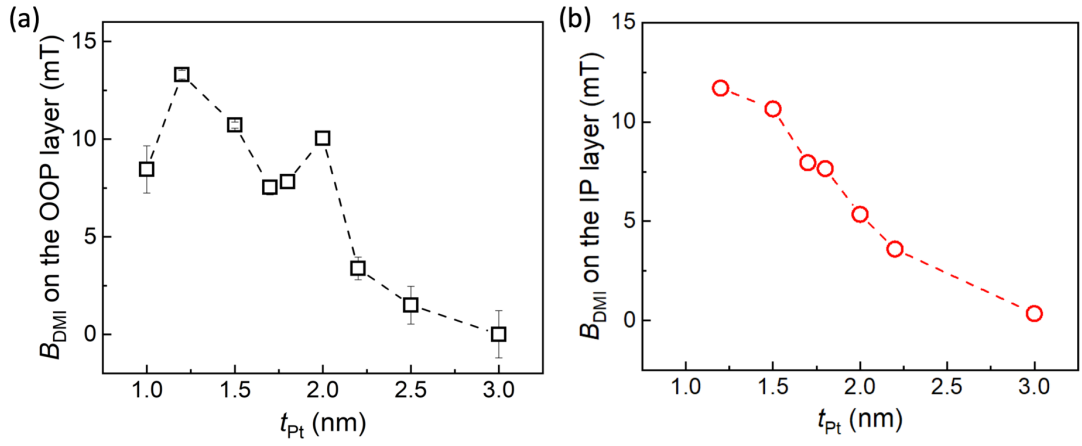

**Figure 4** –Effective $B_{DMI}$ acting on (a) TbFe and (b) Co as a function of Pt thickness. The error bars in (a) represent the standard deviation of the sinusoidal fit to the angular-dependent $\Delta B_c$. The relative errors in (b) are estimated to be about 5% due to uncertainties in determining the position of the magnetoresistance minima.

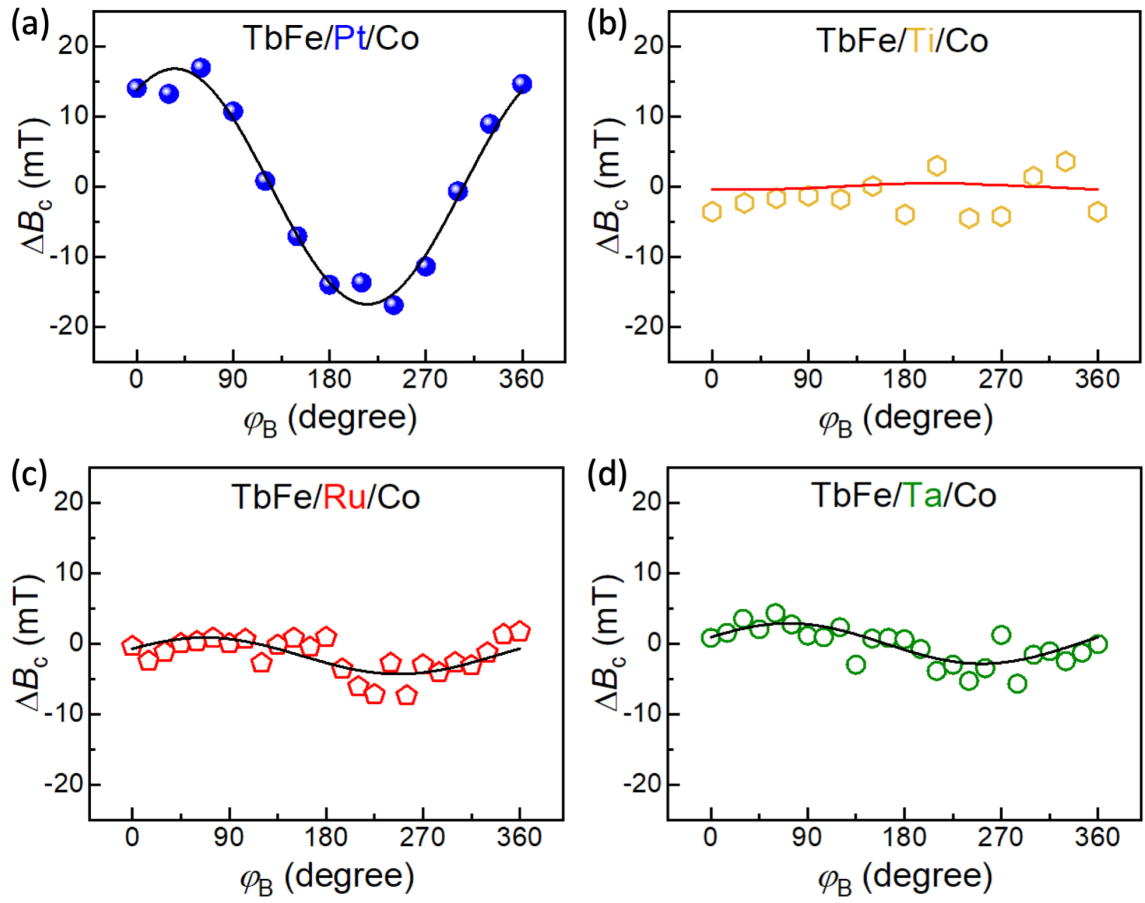

**Figure 5** – (a-d) Coercivity difference $\Delta B_c$ as a function of $\varphi_B$ for samples with Pt, Ti, Ru, and Ta spacers. The spacer thickness was fixed to 1.7 nm while keeping all other layers nominally the same.

SUPPLEMENTAL MATERIAL

**Chiral coupling between magnetic layers with orthogonal magnetization**

Can Onur Avci, Charles-Henri Lambert, Giacomo Sala and Pietro Gambardella

*Department of Materials, ETH Zürich, CH-8093 Zürich, Switzerland*

CONTENTS

## SM 1. Calibration protocol of $\theta_B$

The field sweep and angle scan measurements reported in Fig. 3(a) and 3(c) require an accurate calibration of the angle $\theta_B$ between the z-axis and external field. To perform such a calibration, we apply a large out-of-plane field in excess of the saturation field of the in-plane top Co layer (e.g., 1.5 T). With such field, the magnetizations of both layers are aligned with the external field along the z-axis. We then rotate the field within a small range of angles, e.g., ±10°. Since the anomalous Hall effect signal is maximum at $\theta_B = 0°$, we find the maximum signal and set it as a new reference for $\theta_B = 0°$. We repeat the same procedure by decreasing the range (±5°, ±3°, etc.) and the angle steps progressively until achieving a precise calibration of the angle as shown in Fig. S1.

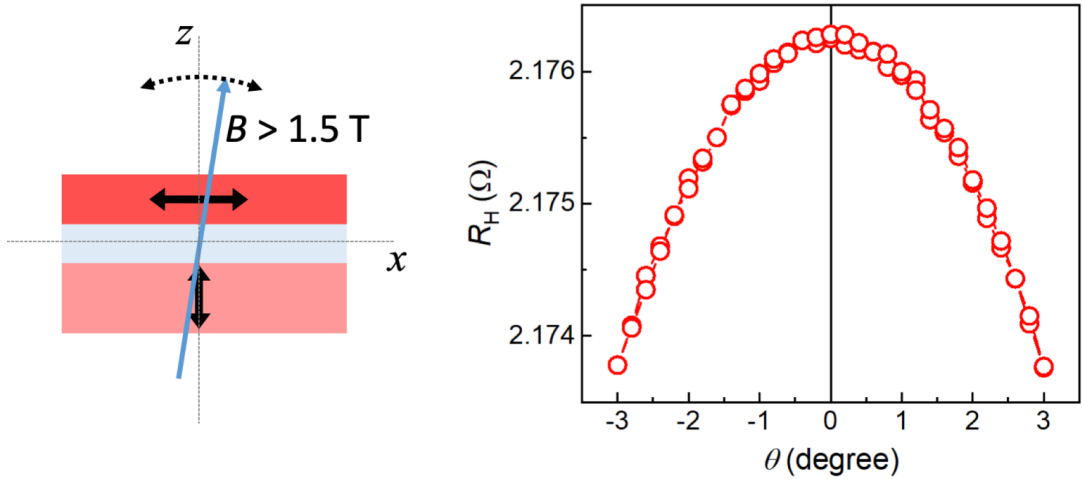

**Figure S1**. Schematic representation of the calibration procedure of $\theta_B$ and representative data of the anomalous Hall resistance in a field of 1.5 T after the calibration.

## SM 2. Measurement of the DMI field as a function of angle

Here we demonstrate an alternative experimental scheme to measure $B_{DMI}$ acting on $\boldsymbol{M}_{TbFe}$ or, more in general, on a layer with OOP magnetization. In this measurement, the magnetic field is rotated about $\boldsymbol{D}$, as shown in Fig. S2(a). The amplitude of the field is set to 200 mT, which is larger than $B_c$ but much lower than the OOP saturation field of $\boldsymbol{M}_{Co}$. In Fig. S2(b) we report $R_H$ for the clockwise (cw) and counterclockwise (ccw) rotation of $\theta_B$ about $\boldsymbol{D}$ measured on TbFe/Pt(1.2 nm)/Co. During the clockwise rotation, we expect that $\boldsymbol{M}_{TbFe}$ reverses from up to down (down to up) upon crossing $\theta_B = 90°$ ($\theta_B = 270°$) when the OOP component of the external field overcomes $B_c$. For the counterclockwise rotation, the sign of the reversal should simply invert. Moreover, in the absence of DMI, the reversal events for clockwise and counterclockwise scans should be symmetric with respect to $\theta_B = 90°$ and 270°. Instead, we observe a clear shift of the reversals towards smaller angles, which indicates that the system switches from an unfavored to a favored configuration when the magnetization rotates in the clockwise direction. This shift agrees with the smaller coercivity observed along $\boldsymbol{D} \times \boldsymbol{z}$ in the field sweep data, as reported in Fig. 3(a). The information obtained from the angle scan measurements is consistent with that derived from the field sweeps discussed in the main text, thus confirming the influence of the interlayer DMI coupling on $\boldsymbol{M}_{\text{TbFe}}$.

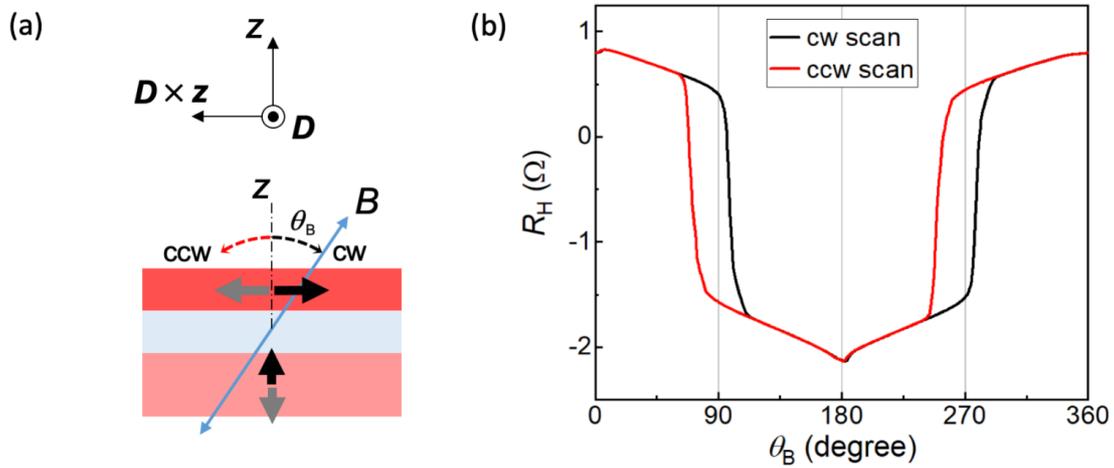

**Figure S2**. (a) Schematic representation of the angle scan in the clockwise (cw) and counterclockwise (ccw) directions. (b) Hall resistance of TbFe(8)/Pt(1.2)/Co(3) measured during the rotation of the magnetic field about $\boldsymbol{D}$.

**SM 3. Macrospin simulations of the influence of the interlayer DMI on magnetization reversal**

We performed macrospin simulations of the interlayer DMI coupling between a perpendicularly magnetized layer (1) and an in-plane layer (2) with a custom-written code using MATLAB. Here, layers 1 and 2 represent the TbFe bottom film and the top Co film, respectively. The total energy of the system comprises the contributions from the Zeeman, magnetic anisotropy, demagnetization, RKKY and DMI energies, and reads:

$$E = -\mathbf{M}_1 \cdot \mathbf{B} - \mathbf{M}_2 \cdot \mathbf{B} - \frac{1}{2}k_1 \sin^2\theta_1 - \frac{1}{2}k_2 \sin^2\theta_2 - \frac{1}{2}\mu_0 M_{S,1}^2 \cos^2\theta_1 - \frac{1}{2}\mu_0 M_{S,2}^2 \cos^2\theta_2 +$$
$$-\frac{1}{2}k_3 \cos^2\varphi_2 - \sigma \mathbf{m}_1 \cdot \mathbf{m}_2 - \mathbf{D} \cdot \mathbf{m}_1 \times \mathbf{m}_2. \quad (S1)$$

Here, $\mathbf{M}_i = M_{S,i}\mathbf{m}_i = M_{S,i}[\sin\theta_i \cos\varphi_i, \sin\theta_i \sin\varphi_i, \cos\theta_i]$ is the magnetization of the $i^{th}$ layer, with saturation magnetization $M_{S,i}$, $k_{1,2}$ are the perpendicular magnetic anisotropy energy densities of the two layers, and $k_3$ is the uniaxial in-plane anisotropy energy density of the 2$^{nd}$ layer. $\sigma$ is the RKKY energy density and $\mathbf{D}$ is the DMI vector, which we assume to be along $-y$.

For each magnetic field magnitude and direction, the equilibrium orientation of $\mathbf{M}_1$ and $\mathbf{M}_2$ was obtained by iteratively minimizing Eq. (S1). The simulations presented here were obtained with the following set of parameters: $M_{S,1} = 0.45 \times 10^6$ A/m, $M_{S,2} = 1.1 \times 10^6$ A/m, $k_1 = 350$ kJ/m$^3$, $k_2 = -50$ kJ/m$^3$, $D = 10$ kJ/m$^3$. For simplicity, we neglected the RKYY coupling, which is zero for orthogonal magnetizations, and the uniaxial in-plane anisotropy ($k_3 = \sigma = 0$). Representative results simulating the field sweeps shown in Fig.3 of the main text and the angle scan in Fig. S2 are presented in Fig. S3. The simulations closely reproduce the measurements and provide a better understanding of the orientation of the two layers. As an example, let us consider a field sweep with the field tilted at $\theta_B = 15°$ along $-\mathbf{D} \times \mathbf{z}$ [initial $\varphi_B = 0°$, black curve in Fig. S3(b)]. At 150 mT, $\mathbf{M}_1$ and $\mathbf{M}_2$ are oriented up-right, a configuration that frustrates the DMI because the latter favors the orientations up-left and down-right (we define the orientation of $\mathbf{M}_2$ by looking at the $xz$ plane from $-y$ to $+y$). When the field is reduced, $\mathbf{M}_2$ rotates towards $-x$ driven by the DMI, which enforces the configuration up-left. Thus, a large field is required to switch $\mathbf{M}_1$ from up to down and bring back the coupled layers in a configuration that opposes the DMI. As the field is increased from -150 mT to 0, $\mathbf{M}_2$ is released and turns to $+x$. Since the new down-right orientation is promoted by the DMI, a large field must be applied to switch $\mathbf{M}_1$ upward, symmetrically to the down-to-up switching. The opposite situation is realized if the field is swept at $\theta_B = 15°$ and tilted along $\boldsymbol{D} \times \boldsymbol{z}$ [initial $\varphi_B = 180°$, black curve in Fig. S3(a)]. In this case, $\boldsymbol{M}_1$ and $\boldsymbol{M}_2$ are initially oriented up-left, which is a configuration favored by the DMI. As the field becomes negative, $\boldsymbol{M}_2$ is forced to rotate towards $x$, which brings the system in a configuration that is unfavored by the DMI. Thus, a smaller field is required to switch $\boldsymbol{M}_1$ from up to down and bring back the coupled layers in a configuration that favors the DMI. A similar situation occurs as $\boldsymbol{M}_1$ switches from down to up as the field returns positive. We can interpret by the same argument the simulations of the angle scan and AMR in Fig. S3(c,d).

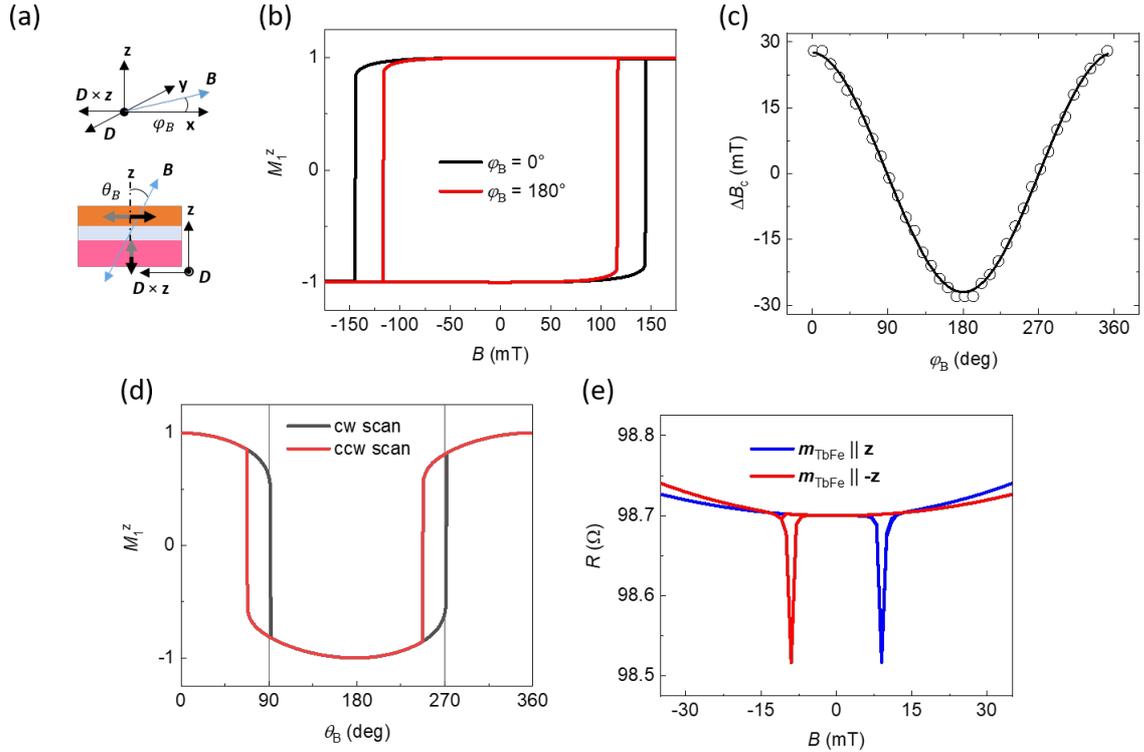

**Figure S3.** (a) Definition of the reference system for field and angle sweep simulations. (b) Macrospin simulation of field sweeps tilted at $\theta_B = 15°$ along $-\boldsymbol{D} \times \boldsymbol{z}$ (initial $\varphi_B = 0°$, black curve) and along $\boldsymbol{D} \times \boldsymbol{z}$ (initial $\varphi_B = 180°$ red curve). The black (gray) arrow shows the orientation of layer 1 (2). (c) Coercivity difference $\Delta B_c$ as a function of $\varphi_B$. (d) Simulation of an angle scan of the external field about $\boldsymbol{D} \parallel -\boldsymbol{y}$. (e) Simulation of the anisotropic magnetoresistance during a field scan at $\theta_B = 89.5°$ and $\varphi_B = 0°$, with $\boldsymbol{M}_1$ oriented along $+z$ (blue curve) and $-z$ (red curve).

**SM 4. Pt Spacer thickness dependence of the TbFe coercivity**

In an effort to understand the deviations in the DMI measured on the TbFe layer (main text, Fig. 4 (a)) we plot the coercivity ($B_c$) of TbFe as a function of the Pt spacer thickness (Fig. S4). As evident from the data, the samples with Pt(1 nm) and Pt(2 nm) spacer show significant deviations from the overall linearly increasing $B_c$ trend as a function of $t_{Pt}$. Such deviations suggest that the properties of these TbFe layers might be different than the remaining samples in the batch. The magnetic properties as well as the interlayer DMI is known to be highly sensitive to the interfaces of the ferromagnets with Pt, which might exhibit local fluctuations due to the influence of substrate, sputtering process and fabrication-related issues. We believe that one or several of such factors played a role for the mentioned two samples and resulted in a different coercivity and interfacial DMI simultaneously.

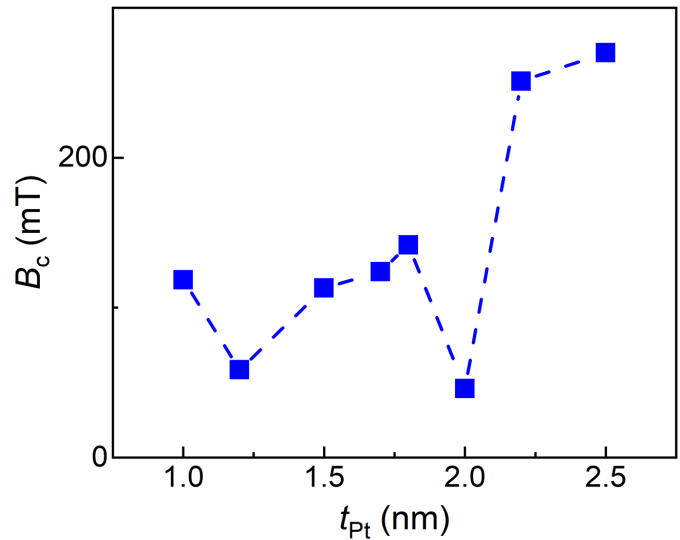

**Figure S4.** Coercivity ($B_c$) of the TbFe layer as a function of the Pt spacer thickness.